\documentclass[envcountsame,fleqn]{llncs}

\usepackage{amsmath}
\usepackage{amsfonts}
\usepackage{pslatex}
\usepackage{graphicx}
\usepackage{color}
\usepackage{url}

\allowdisplaybreaks

\title{%
  Event Systems and Access Control\thanks{%
    This work was partly supported by the project Desirs of
    ACI S\'ecurit\'e Informatique.}
}
\author{%
  Dominique M\'ery \and Stephan Merz\\
  \email{\{Dominique.Mery,Stephan.Merz\}@loria.fr}
}
\authorrunning{D.~M\'ery and S.~Merz}
\institute{%
  LORIA -- INRIA Lorraine -- Universit\'e Henri Poincar\'e\\
  Nancy, France
}

\renewcommand{\O}[1]{\overline{#1}}
\newcommand{\fun}{\rightarrow}
\newcommand{\deq}{\mathrel{\stackrel{\scriptscriptstyle\Delta}{=}}}
\newcommand{\nat}{\mathbb{N}}
\newcommand{\rat}{\mathbb{Q}}

\newcommand{\kw}[1]{\textbf{#1}}
\newcommand{\event}{\kw{event}}
\newcommand{\fairness}{\kw{fairness}}
\newcommand{\permission}{\kw{permission}}
\newcommand{\prohibition}{\kw{prohibition}}
\newcommand{\obligation}{\kw{obligation}}
\newcommand{\allow}{\kw{right}}

\newcommand{\fis}{\mathop{\textbf{fis}}}
\newcommand{\stable}{\mathop{\textbf{stable}}}
\newcommand{\inv}{\mathop{\textbf{inv}}}

\newcommand{\false}{\textbf{false}}
\renewcommand{\implies}{\Rightarrow}

\newenvironment{noj}{\begin{array}[t]{@{}l@{}}}{\end{array}}
\newenvironment{conj}{\begin{array}[t]{@{\mbox{$\land\ $}}l@{}}}{\end{array}}
\newenvironment{disj}{\begin{array}[t]{@{\mbox{$\lor\ $}}l@{}}}{\end{array}}

\newcommand{\Rule}[3]{%
  \begin{array}[b]{@{}c@{}l@{}}
    \begin{aligned}
      #2
    \end{aligned}\\
    \cline{1-1} & \raisebox{1.2ex}{\ \ \small\smash{\text{#1}}}\\[-2ex]
    \begin{array}{@{}c@{}}
      #3\!\!\!
    \end{array}
  \end{array}}

\makeatletter
\newcounter{abr@ctr}
\newcommand{\abr@c}{\c@abr@ctr\advance\c@abr@ctr\@ne}

\ifx\documentclass\undefined
\else
  \DeclareSymbolFont{tlaitalics}{\encodingdefault}{ptm}{m}{it}
  \let\itfam\symtlaitalics
\fi

\newcommand{\noTeXmath}{%
\c@abr@ctr=\itfam
\multiply\c@abr@ctr"100\relax
\advance\c@abr@ctr "7061\relax
\mathcode`a=\abr@c
\mathcode`b=\abr@c
\mathcode`c=\abr@c
\mathcode`d=\abr@c
\mathcode`e=\abr@c
\mathcode`f=\abr@c
\mathcode`g=\abr@c
\mathcode`h=\abr@c
\mathcode`i=\abr@c
\mathcode`j=\abr@c
\mathcode`k=\abr@c
\mathcode`l=\abr@c
\mathcode`m=\abr@c
\mathcode`n=\abr@c
\mathcode`o=\abr@c
\mathcode`p=\abr@c
\mathcode`q=\abr@c
\mathcode`r=\abr@c
\mathcode`s=\abr@c
\mathcode`t=\abr@c
\mathcode`u=\abr@c
\mathcode`v=\abr@c
\mathcode`w=\abr@c
\mathcode`x=\abr@c
\mathcode`y=\abr@c
\mathcode`z=\abr@c
\c@abr@ctr=\itfam
\multiply\c@abr@ctr"100\relax
\advance\c@abr@ctr "7041\relax
\mathcode`A=\abr@c
\mathcode`B=\abr@c
\mathcode`C=\abr@c
\mathcode`D=\abr@c
\mathcode`E=\abr@c
\mathcode`F=\abr@c
\mathcode`G=\abr@c
\mathcode`H=\abr@c
\mathcode`I=\abr@c
\mathcode`J=\abr@c
\mathcode`K=\abr@c
\mathcode`L=\abr@c
\mathcode`M=\abr@c
\mathcode`N=\abr@c
\mathcode`O=\abr@c
\mathcode`P=\abr@c
\mathcode`Q=\abr@c
\mathcode`R=\abr@c
\mathcode`S=\abr@c
\mathcode`T=\abr@c
\mathcode`U=\abr@c
\mathcode`V=\abr@c
\mathcode`W=\abr@c
\mathcode`X=\abr@c
\mathcode`Y=\abr@c
\mathcode`Z=\abr@c}

\makeatother
\noTeXmath

\begin{document}

\maketitle

\begin{abstract}
  We consider the interpretations of notions of access control
  (permissions, interdictions, obligations, and user rights) as
  run-time properties of information systems specified as event
  systems with fairness. We give proof rules for verifying that an
  access control policy is enforced in a system, and consider
  preservation of access control by refinement of event systems. In
  particular, refinement of user rights is non-trivial; we propose to
  combine low-level user rights and system obligations to implement
  high-level user rights.
\end{abstract}

\section{Introduction}
\label{sec:introduction}

The specification of access control policies for information systems
is a fundamental building block of a methodology for describing and
assessing the security of information infrastructure. Existing
languages for describing access control such as
RBAC~\cite{sandhu:rbac} and OrBAC~\cite{abou:orbac} focus on the
static structure of information systems. They identify the actors
(abstractly represented as roles), objects (abstracted as views), and
activities that intervene in an information system, and then impose
constraints on activities, in the form of permissions and
prohibitions. Certain formalisms also encompass more advanced security
properties such as rights or obligations of actors to perform certain
activities. OrBAC makes a step toward specifying access control
policies that may depend on run-time information by associating rights
with contexts. However, it is not possible within OrBAC to verify that
a system enforces a given access control policy, because the dynamic
behavior of the system is not modeled.

In this paper, we propose to relate the specification of access
control policies to formal models of dynamic system behavior, and we
give proof rules to demonstrate that a system implements an access
control policy. Changing somewhat the perspective, one can also pose
the problem of deriving a security monitor that enforces a policy for
a fixed, underlying system.

We describe information systems within the well-known paradigm of
event systems, see
e.g.~\cite{abrial:b-book,abrial:event,back:refinement}. Run-time
properties of event systems can be specified as formulas of temporal
logic, and there are well-established verification rules to derive
properties of event systems. We are therefore led to interpret access
control primitives as properties of runs of event systems: permissions
and prohibitions are easily expressed as constraints on the enabling
condition of events. Dually, the right of an actor to perform a
certain activity can be expressed as a lower bound on the enabling
condition. The interpretation of obligations is less obvious, and we
propose to interpret them as liveness properties, expressible in
temporal logic.

Event systems have traditionally been associated with a formal
development method based on stepwise refinement. We therefore consider
how access control annotations are preserved under system
refinement. Because permissions, prohibitions, and obligations are
interpreted as safety and liveness properties of runs, standard
results about refinement of event systems ensure that they are
preserved by refinement.  Preservation of user rights requires extra
conditions, and the precise formulation is non-trivial when the
``grain of atomicity'' of a system description changes during
refinement. We propose a condition that relies on a combination of
concrete-level user rights and obligations. We illustrate our
proposals with a running example of a simple loan management system on
which different access control requirements can be imposed.

\paragraph{Related work.}

The existing literature on formalisms for the specification of access
control considers mainly static methods of analysis. For example,
Bertino et al.~\cite{bertino:logical} and Cuppens et
al.~\cite{cuppens:misconfiguration}, among others, analyze security
policies for inconsistencies, and Benferhat et
al.~\cite{benferhat:stratification} consider techniques to resolve
such inconsistencies based on stratification of rules.

Closer to our concerns is work by Ryan et
al.~\cite{guelev:access-control,zhang:evaluating} on the use of model
checking for verifying access-control policies. However, we work in a
deductive framework, and we are mainly interested in verifying
refinement relationships. Koch et al.~\cite{koch:access-control}
suggest a UML notation for specifying access control, together with a
semantics based on graph transformation and corresponding analysis
techniques. More distantly related is the work around
UMLSec~\cite{juerjens:umlsec}, which is mainly concerned with secrecy
properties. In particular, J\"urjens~\cite{juerjens:secrecy} considers
the preservation of secrecy properties by the refinement concepts of
the specification language Focus.

\section{Fair Event Systems}
\label{sec:event-systems}

We use the well-known paradigm of event
systems~\cite{abrial:b-book,abrial:event,back:refinement}, extended by
weak fairness conditions, to express system models.\footnote{%
  Adding strong fairness would not pose any conceptual problems, but
  it would complicate the presentation because we would have to
  introduce more elaborate temporal logic operators.}
This section gives a brief overview over the syntax we use to describe
systems and their properties, and introduces associated verification
rules.

\subsection{Event systems and their runs}
\label{sec:syntax-fair-event}

A system specification lists the constant parameters, including any
underlying sets, functions, and relations that describe the data over
which the system operates. A \emph{constant assumption} $Hyp$
constrains the values of these constants; it is syntactically
expressed as a first-order logic formula over the constant parameters.

More importantly, a specification declares a tuple $var$ of state
variables that represent the current state of the system. The runs of
a system are characterized by an \emph{initial condition}, which is a
state predicate $Init$ over the variables $var$, and a list of
\emph{events} that describe the possible system transitions. We write
the definition of an event $e$, with list of parameters $x$, as
\[
\begin{noj}
  \event\ e(x) = BA_e(x,var,var')\\
  \fairness\ fair_e(x,var)
\end{noj}
\]
In such a definition, $BA_e$ is the before-after predicate for the
event $e$; this is a first-order formula built from the constants
declared for the system specification, the event's parameters $x$, as
well as primed and unprimed occurrences of the system variables
$var$. As is conventional, a primed occurrence $v'$ of a state
variable $v$ denotes the value of $v$ in the state following the
transition described by $BA_e$, while an unprimed occurrence denotes
the value of $v$ in the state before the transition.  Each event is
associated with a fairness condition, expressed by a predicate
$fair_e(x,var)$.  Intuitively, the fairness condition rules out traces
where the predicate $fair_e(x,var)$ remains true but the event $e(x)$
never occurs.

For an event $e(x)$, we define its feasibility condition
\begin{equation}
  \label{eq:fis}
  \fis e(x)\ \deq\ \exists var': BA_e(x,var,var')
\end{equation}
by existentially quantifying over the primed occurrences of the state
variables; thus, the state predicate $\fis e(x)$ is true of those
states that have a successor state related by an occurrence of the
event $e(x)$.

Finally, a system specification should provide an \emph{invariant}
that constrains the set of reachable states, syntactically specified
by a state predicate $Inv$ over the system variables $var$.

A system specification is well-formed if all of the following conditions
hold:
\begin{itemize}
\item The initial condition implies the invariant:
  \begin{equation}\label{eq:init-inv}
    Hyp\ \models\ Init(var) \implies Inv(var)
  \end{equation}
\item The invariant is preserved by any event $e$, for any
  instantiation of the parameters:
  \begin{equation}\label{eq:event-inv}
    Hyp\ \models\ Inv(var) \land BA_e(x,var,var') \implies Inv(var')
  \end{equation}
\item For any event, the fairness condition implies the feasibility of
  the event:
  \begin{equation}\label{eq:fair-feasible}
    Hyp\ \models\ Inv(var) \land fair_e(x,var) \implies \fis e(x)
  \end{equation}
  Observe that we allow the fairness condition to be strictly stronger
  than the feasibility predicate. For example, an event without
  fairness assumption can be modeled by declaring the fairness
  condition to be $\false$.
\end{itemize}

In the following, we simplify the notation by writing $P$ and $P'$ for
$P(var)$ and $P(var')$ when $P$ is a state predicate and $A(x)$ for
$A(x,var,var')$ when $A$ is a formula that contains both primed and
unprimed occurrences of state variables, such as a before-after
predicate.

\begin{figure}[tp]
  \centering
  \begin{tabular}{l}
    \kw{system} $Bank$\\
    \quad\kw{constants} $Client$, $Loan$, $maxDebt$\\
    \quad\kw{assumption} $Client \neq \emptyset
       \land Loan \neq \emptyset
       \land maxDebt \in \rat$\\
    \quad\kw{variables} $clt, loans, due, rate, maxExtra, extra$\\[1mm]
    \quad\kw{invariant}
    $\begin{conj}
      loans \subseteq Loan\\
      clt \in [loans \fun Client]
      \land due \in [loans \fun \rat]
      \land rate \in [loans \fun \rat]\\
      maxExtra \in [loans \fun \rat]
      \land extra \in [loans \fun \rat]\\
      \forall c \in Client: 
        \big(\sum \{due(ll) : ll \in loans \land clt(ll)=c\}\big) \leq maxDebt
    \end{conj}$\\[1mm]
    \quad\kw{initial} 
      $loans = \emptyset
       \land clt = \emptyset
       \land due = \emptyset
       \land rate = \emptyset
       \land maxExtra = \emptyset
       \land extra = \emptyset$\\[1mm]
    \quad\kw{event} $newLoan(c, l, amt, dur, mx) =$\\
    \qquad$\begin{conj}
      c \in Client \land l \in Loan \setminus loans \land amt \in \rat \land dur \in \nat\\
      amt + \big(\sum \{due(ll) : ll \in loans \land clt(ll)=c\}\big) \leq maxDebt\\
      loans' = loans \cup \{l\}
      \land clt' = clt \cup \{l \mapsto c\}\\
      due' = due \cup \{l \mapsto sum\}
      \land rate' = rate \cup \{l \mapsto sum / dur\}\\
      maxExtra' = maxExtra \cup \{l \mapsto mx\}
      \land extra' = extra \cup \{l \mapsto 0\}
    \end{conj}$\\
    \quad\kw{fairness} $\false$\\[1mm]
    \quad\kw{event} $payRate(l) =$\\
    \qquad$\begin{conj}
      l \in loans\\
      due' = due \oplus \{l \mapsto due(l) - rate(l)\}\\
      loans' = loans \land clt' = clt \land rate' = rate
      \land maxExtra' = maxExtra \land extra' = extra
    \end{conj}$\\
    \quad\kw{fairness} $l \in loans \land due(l) > 0$\\[1mm]
    \quad\kw{event} $extraPayBack(l, amt) =$\\
    \qquad$\begin{conj}
      l \in loans \land amt \in \rat\\
      due' = due \oplus \{l \mapsto due(l) - amt\}
      \land extra' = extra \oplus \{l \mapsto extra(l) + amt\}\\
      loans' = loans \land clt' = clt \land rate' = rate
      \land maxExtra' = maxExtra
    \end{conj}$\\
    \quad\kw{fairness} $\false$\\
    \kw{end system}
  \end{tabular}
  \caption{Sample system specification.}
  \label{fig:bank}
\end{figure}

Figure~\ref{fig:bank} shows a specification of a simple event system
that will serve as a running example for this paper\footnote{%
  We adopt the convention of writing long conjunctions and
  disjunctions as ``lists'' bulleted with $\land$ and $\lor$, relying
  on indentation to save parentheses.
}.
It models a simple management system for loans: clients can take out
loans provided they are not overly indebted, and they should pay them
back, either by paying the rates due or via extra payments. The
specification is written in a language of set theory where functions
are sets of pairs $x \mapsto y$ and where $\oplus$ denotes function
override.  It is easy to verify that this specification is well-formed
according to the above criteria. At this point, we only give the model
of the base information system, it will later be extended with
annotations corresponding to access control primitives.

Runs of a system specification are $\omega$-sequences $\sigma = s_0
s_1 \ldots$ of states (i.e., valuations of variables) that satisfy the
following conditions:
\begin{itemize}
\item the initial state $s_0$ satisfies the initial condition,
\item any two successive states $(s_i,s_{i+1})$ either satisfy the
  before-after predicate $BA_e(x)$ for some event $e$ and some
  parameter values $x$, or agree on the values of all system variables
  $var$ (so-called stuttering steps), and
\item $\sigma$ satisfies all fairness conditions: for each event $e$
  and all parameter values $x$ there are infinitely many positions $i
  \in \nat$ such that either the fairness condition $fair_e(x)$ is
  false at $s_i$ or $(s_i,s_{i+1})$ satisfy $BA_e(x)$.
\end{itemize}

The well-formedness conditions (\ref{eq:init-inv}) and
(\ref{eq:event-inv}) above ensure that each state $s_i$ of a system
run satisfies the system invariant. If only countably many event
instances are feasible at each state of a system run, the condition
(\ref{eq:fair-feasible}) implies that the specification is
machine-closed~\cite{abadi:existence}, but this observation will not
play a role in the remainder of this paper.

\subsection{Properties of runs}
\label{sec:properties}

\begin{figure}[tp]
  \centering
  \(\begin{array}{c}
    \Rule{(stable)}{%
      P \land BA_e(x) \implies P'\quad\text{for all events $e(x)$}}{%
      \stable P}
    \\[2mm]
    \Rule{(induct)}{%
      Init \implies P \quad \stable P}{%
      \inv P}
    \qquad
    \Rule{(inv-weaken)}{%
      \inv P \quad P \implies Q}{%
      \inv Q}
    \\[2mm]
    \Rule{(fair)}{%
      \begin{array}{c}
        P \land BA_a(x) \land \lnot BA_e(t) \implies P' \lor Q'
        \quad\text{for all events $a(x)$}\\
        P \implies fair_e(t)
      \end{array}}{%
      P \leadsto Q \lor (P \land e(t))}
    \\[2mm]
    \Rule{(wfo)}{%
      \forall x \in S: F(x) \leadsto G \lor (\exists y \in S: y \prec x \land F(y))
      \quad (S,\prec)\ \text{well-founded}}{%
      (\exists x \in S: F(x)) \leadsto G}
    \\[2mm]
    \Rule{(effect)}{%
      P \land BA_e(t) \implies Q'}{%
      P \land e(t) \leadsto Q}
    \qquad
    \Rule{(refl)}{%
      F \implies G}{%
      F \leadsto G}
    \qquad
    \Rule{(inv-leadsto)}{%
      \inv I \quad I \land F \leadsto G}{%
      F \leadsto G \land I}
    \\[2mm]
    \Rule{(trans)}{%
      F \leadsto G \quad G \leadsto H}{%
      F \leadsto H}
    \qquad
    \Rule{(disj)}{%
      F \leadsto H \quad G \leadsto H}{%
      F \lor G \leadsto H}
    \qquad
    \Rule{(exists)}{%
      F(x) \leadsto G(x)}{%
      (\exists x:F(x)) \leadsto (\exists x:G(x))}
  \end{array}\)
  \caption{Verification rules for fair event systems.}
  \label{fig:rules}
\end{figure}

We can reason about the runs of fair event systems using elementary
temporal logic. For the purposes of this paper, we consider safety
properties $\stable P$ and $\inv P$ where $P$ is a state predicate,
and liveness properties $F \leadsto G$ (``$F$ leads to $G$'') where
$F$ and $G$ are Boolean combinations of state predicates and event
formulas $e(x)$ for events $e$ of the underlying event system. These
formulas are interpreted over a run $\sigma = s_0 s_1 \ldots$ as
follows:
\[
\begin{array}{@{}l@{\quad\text{iff}\quad}l@{}}
  \sigma \models \stable P &
  \text{for all $n \in \nat$, if $\sigma|_n \models P$ then 
    $\sigma|_m \models P$ for all $m \geq n$}\\
  \sigma \models \inv P &
  \text{$\sigma|_n \models P$ for all $n \in \nat$}\\
  \sigma \models F \leadsto G &
  \text{for all $n \in \nat$, if $\sigma|_n \models F$ then
    $\sigma|_m \models G$ for some $m \geq n$}
\end{array}
\]
In these definitions, $\sigma|_n \models F$ means that formula $F$
holds of the suffix of $\sigma$ from point $n$ onwards: if $F$ is a
state predicate then it should be satisfied at state $s_n$, if $F$ is
an event formula $e(x)$ then the defining action formula $BA_e(x)$
should hold of the pair of states $(s_n, s_{n+1})$.

Figure~\ref{fig:rules} contains proof rules for deriving properties of
fair event systems; similar proof rules can be found, for example, in
papers on the Unity~\cite{misra:logic} or TLA~\cite{lamport:tla}
formalisms. As before, $Init$ denotes the initial condition of the
system specification, $BA_e(t)$ denotes the before-after predicate
defining the event instance $e(t)$, and $fair_e(t)$ represents the
fairness condition associated with that event instance. The variable
$x$ in rules (stable) and (fair) is assumed to be different from the
free variables of $P$, $Q$ or $BA_e(t)$.

The rule (fair) is the basic proof rule for establishing leadsto
properties; its soundness relies on the underlying assumption of weak
fairness. Rule (wfo) allows us to derive liveness properties by
induction over some well-founded ordering. The remaining rules can be
used to combine elementary leadsto formulas. In proving the
non-temporal hypotheses of these rules, we may of course use any
assumptions on the constants appearing in a system specification.

\section{Specifying Access Control}
\label{sec:access-control}

Access control policies describe the conditions under which events may
occur. Typically, one first specifies the actors (roles), objects
(views), and activities of an information system, and then describes
which actors are allowed to (or not allowed to) perform which
activities on which objects. The OrBAC formalism~\cite{abou:orbac}
refines this general idea: first, access control policies are
described within organisations (e.g., a hospital or a bank). Second,
and more significantly, one can specify conditions under which an
access is allowed by defining a ``context'' of access. Moreover,
roles, views, and activities are arranged in hierarchies, with access
rules for instances systematically derived with the help of
inheritance rules~\cite{cuppens:inheritance}.

OrBAC thus provides a declarative, PROLOG-like language to define
access control policies. The dynamic aspect of a system is captured by
the notion of context, which can be defined in terms of the system
state. It is straightforward to translate an OrBAC model into an event
system: the static structure of roles and views is represented by the
constant parameters of a system, activities correspond to the system's
events, and contexts are defined as state predicates. Without
completely formalizing this translation, we now consider how event
systems can be extended to describe access control
policies.\footnote{%
  The running example of Fig.~\ref{fig:bank} does not mention roles
  (actors), but it should be obvious how to include them in the static
  model.}
The interest in doing so can be twofold: first, an event system can be
developed in order to verify that it satisfies a given policy. Second,
one may be interested in enforcing an access control policy over a
fixed underlying system by imposing a security monitor.  We will
consider both of these views for different access control primitives:
permissions and prohibitions, user rights, and obligations.

\subsection{Permissions and prohibitions}
\label{sec:permission}

At its base, an access control policy describes when an activity is
\emph{permitted}, and when it is \emph{forbidden}. Whereas permissions
and prohibitions should be mutually exclusive, they need not cover all
possible situations in cases where the policy is not completely
specified.

We represent permissions and prohibitions by associating two more
predicates (besides the fairness predicate already introduced in
Sect.~\ref{sec:syntax-fair-event}) with event definitions. For
example, a security policy might specify
\[
\begin{noj}
  \event\ newLoan(c,l,amt,dur,mx)\\
  \permission\ l \notin loans \land risk(c,amt) \in \{low,medium\}
     \land mx \leq maxPayback(amt,dur)\\
  \prohibition\ risk(c,amt) = high
\end{noj}
\]
to indicate that a new loan for a client may be approved if the
associated risk (evaluated according to some unspecified risk
function) is below a certain threshold value and if the maximum amount
permitted for extra payback is within certain bounds, and that a new
loan must not be approved if the risk is too high.

An event system implements the permissions and prohibitions declared
in a security policy if the event is feasible only if it is permitted
and infeasible when it is forbidden. Formally, we obtain the proof
obligations
\begin{eqnarray}
  \label{eq:permission}
  Hyp & \models & Inv \land \fis e(x) \implies perm_e(x)\\
  \label{eq:prohibition}
  Hyp & \models & Inv \land proh_e(x) \implies \lnot \fis e(x)
\end{eqnarray}
where $perm_e(x)$ and $proh_e(x)$ are the permission and prohibition
predicates associated with event $e$, and $Inv$ and $Hyp$ are the
system invariant and the constant assumptions, as before.

The event system of Fig.~\ref{fig:bank} does not implement the above
permissions and prohibitions, as it does not evaluate the risk
associated with a loan. A simple way of ensuring that a system
implements the permission and prohibition clauses of a security policy
is to conjoin $perm_e(x) \land \lnot proh_e(x)$ to the before-after
predicate of the event definition. Alternatively, the access control
policy can be ensured at run time by a separate monitor that allows
events to be activated only if the permissions and prohibitions are
respected.

Observe, however, that strengthening the guard of an event may
invalidate the well-formedness condition (\ref{eq:fair-feasible}) that
states that the fairness predicate of an event should imply its
feasibility. We therefore add the following proof obligation to the
well-formedness conditions of an event system with permissions and
prohibitions:
\begin{equation}
  \label{eq:fair-perm-proh}
  Hyp\ \models\ Inv \land fair_e(x) \implies perm_e(x) \land \lnot proh_e(x).
\end{equation}
This condition is trivially satisfied for the event $newLoan$ of our
running example, because no fairness is required of that event.

\subsection{User rights}
\label{sec:user-rights}

Permissions and prohibitions restrict the feasibility of
events. Dually, it may be interesting to specify \emph{user rights}:
conditions that spell out when an activity should be permitted in a
system. User rights can again be represented in event systems by
associating a corresponding predicate with an event. For example, we
may wish to state explicitly that a client has the right to make extra
payments within the agreed-upon limits:
\[
\begin{noj}
  \event\ extraPayBack(l,amt)\\
  \allow\ l \in loans \land amt \in \rat \land amt + extra(l) \leq maxExtra(l).
\end{noj}
\]

An event system implements a user right if the event is feasible
whenever the predicate specifying the right holds:
\begin{equation}
  \label{eq:right}
  Hyp\ \models\ Inv \land right_e(x) \implies \fis e(x)
\end{equation}
Because a security monitor can only schedule existing events of the
underlying system, user rights will have to be verified over the event
system itself rather than enforced by a monitor. However, the monitor
will have to observe a similar condition to make sure that an event
permitted by a right is never disabled by the monitor.

The conditions (\ref{eq:permission}), (\ref{eq:prohibition}), and
(\ref{eq:right}) show that the right to perform an activity should
imply (assuming the system invariant) that the activity is allowed,
and that it is not forbidden. It is not unreasonable for a user right
to be strictly stronger than the corresponding permission, or than the
feasibility of the event. For example, a bank may accept extra
payments beyond the pre-determined bound at its discretion.

User rights can be understood as branching-time properties: whenever
the predicate $right_e(t)$ is true, the system has a possible
continuation that begins with the event (instance) $e(t)$, and we will
take up this discussion in Sect.~\ref{sec:refine-secure}.

\subsection{Obligations}
\label{sec:obligations}

Languages for access control policies such as OrBAC also include
primitives for specifying \emph{obligations}. Intuitively, whereas a
user right states when a certain activity \emph{may} occur, an
obligation asserts that the activity \emph{should} occur. The
article~\cite{abou:orbac} introducing the OrBAC notation does not
define a formal semantics for obligations, but concepts of permission,
rights, and obligations have traditionally been the domain of deontic
logic~\cite{hilpinen:deontic,meyer:deontic}. To our knowledge, the
interpretation of formulas of deontic logic over models of information
systems such as event systems has not been studied, and we do not wish
to introduce this extra complication.

As before, we associate obligations with events by defining
suitable predicates. In our running example, we might want to assert
that a user has an obligation to pay the rates as long as they are due
by writing
\[
\begin{noj}
  \event\ payRate(l)\\
  \obligation\ l \in loans \land due(l)>0.
\end{noj}
\]

What does it formally mean for an event system to implement an
obligation?  A first idea would be to interpret an obligation to
perform a certain activity as prohibiting the system from performing
any other activity. However, this interpretation appears to be
unreasonably stringent and prone to contradictions. For example, a
user of a computer system may have an obligation to regularly change
his password, but he can do so only when logged in. Clearly, the
obligation to change the password should not preclude the user from
logging in, although it is conceivable that one could then prevent the
user from doing anything but changing his password.

We believe instead that obligations can, in a first approximation, be
interpreted as liveness properties, and can be formalized in temporal
logic. The two following interpretations appear particularly plausible.
\begin{eqnarray}
  \label{eq:strict-obl}
  \text{strict obligation:} && obl_e(x)\ \leadsto\ e(x)\\
  \label{eq:weak-obl}
  \text{weak obligation:} && obl_e(x)\ \leadsto\ \lnot obl_e(x) \lor e(x)
\end{eqnarray}

The strict interpretation of obligations requires that the event
occurs eventually whenever the obligation arises. Under the weak
interpretation, the obligation ceases as soon as the predicate $obl_e$
becomes false, which need not be due to an occurrence of $e$.  In our
example, the weak interpretation appears more reasonable: it is
satisfied when a client pays back the loan via an extra payment.
Observe that the weak interpretation of an obligation coincides with
the interpretation of a weak fairness requirement, with $obl_e(x)$ as
the fairness condition.

Whatever interpretation is chosen, the proof rules of
Fig.~\ref{fig:rules} can be used to verify that a fair event system
implements its obligations. The temporal interpretation of obligations
may also be of interest when one is interested in deriving a security
monitor that enforces obligations for a given system, at least for
controllable events. To do so, one could apply recent work on
controller synthesis based on game-theoretic
interpretations~\cite{arnold:games-partial}, but we do not pursue this
idea any further here.

In some applications, the interpretation of obligations as liveness
properties may be too abstract, and it would be more natural to
indicate real-time deadlines for obligations (``the payment should
be received before the end of the current month''). We do not consider
real-time specifications in this paper.

\section{Refinement of System Specifications}
\label{sec:refinement}

Stepwise methods of system development insist that systems should be
developed in a succession of models that gradually add representation
detail and that introduce new correctness properties. The key
requirement for a sensible notion of refinement is that system
properties that have been established at higher levels of abstraction
are preserved by construction so that they do not have to be
reproven. Refinement-based approaches help to discover potential
problems early on. They also distribute the overhead of formal
verification over the entire development process.  We will first
consider verification conditions for proving refinement of fair event
systems that preserve temporal logic properties. In a second step, we
will study how refinement interacts with the access control primitives
considered in Sect.~\ref{sec:access-control}.

\subsection{Refinement of fair event systems}
\label{sec:refine-event}

Standard refinement notions for event systems are known to preserve
safety properties, and extensions for liveness and fairness properties
have also been considered, for example
in~\cite{abrial:dynamic,back:refinement-fair}. In the following, we
make use of the language of temporal logic of
Sect.~\ref{sec:properties} to state verification conditions for
preserving liveness properties at a higher level of abstraction than
in traditional formulations.

Refined models describe the system at a finer level of granularity and
typically introduce new events that have no observable effect at the
previous levels of abstraction.  Formally, we assume (without loss of
generality) that the refinement is described with the help of a tuple
$var_{ref}$ of variables disjoint from the variables $var_{abs}$ used
in the original model. The two state spaces are related by a
\emph{gluing invariant} $J$, a state predicate built from the
variables $var_{abs}$ and $var_{ref}$, and the constant parameters of
both models. We may assume that $J$ implies both the abstract-level
and the concrete-level invariants $Inv_{abs}$ and $Inv_{ref}$.  An
event $ea(x)$ of the abstract model may be refined by a number of
low-level events $er_1(x,y_1)$, \ldots, $er_n(x,y_n)$; for technical
simplicity, we assume that all parameters of $ea$ are also parameters
of $er_i$, although this assumption could easily be removed.  Also,
new events $en(z)$ may be introduced in the refined model.

An event system $Ref$ is a refinement of an event system $Abs$
with respect to the gluing invariant $J$ if $Ref$ is itself
well-formed according to the conditions (\ref{eq:init-inv}),
(\ref{eq:event-inv}), and (\ref{eq:fair-feasible}), and if moreover
all the following conditions hold (again, we drop the variables that
occur in the respective predicates; besides, $Hyp$ denotes the
conjunction of the abstract- and concrete-level constant assumptions).
\begin{itemize}
\item Every initial state of the refinement can be mapped to a
  corresponding initial state of the abstract specification:
  \begin{equation}
    \label{eq:ref-init}
    Hyp\ \models\ Init_{ref} \implies \exists var_{abs}: Init_{abs} \land J
  \end{equation}
\item Events of the refinement can be mapped to events or to
  stuttering transitions of the abstract specification. There are two
  cases:
  \begin{itemize}
  \item If event $er(x,y)$ refines an abstract event $ea(x)$ then its
    effect can be mapped to an occurrence of $ea$:
    \begin{equation}
      \label{eq:ref-event}
      Hyp\ \models\ J \land BA_{er}(x,y)
      \implies \exists var_{abs}': BA_{ea}(x) \land J'
    \end{equation}
  \item If event $en(z)$ is a new event then its effect is invisible
    at the abstract level\footnote{%
      As suggested in~\cite{abrial:refinement}, this requirement could
      be weakened by requiring that event $en(z)$ merely preserves the
      high-level invariant.}:
    \begin{equation}
      \label{eq:ref-skip}
      Hyp\ \models\ J \land BA_{en}(z)
      \implies \exists var_{abs}': var_{abs}'=var_{abs} \land J'
    \end{equation}
  \end{itemize}
\item The refinement preserves the fairness constraints of the
  abstract level. Formally, assume that the abstract event $ea(x)$ is
  refined by low-level events $er_1(x,y_1)$, \ldots, $er_n(x,y_n)$:
  \begin{equation}
    \label{eq:ref-fair}
    Ref\ \models\ 
    \begin{noj}
      J \land fair_{ea}(x)\\
      \quad\leadsto\ 
      ea_1(x) \lor \ldots \lor ea_n(x)
      \lor \lnot\exists var_{abs}: J \land fair_{ea}(x)
    \end{noj}
  \end{equation}
  where the ``abstract trace'' $ea_i(x)$ of $er_i(x,y_i)$ is defined
  as
  \[
    ea_i(x)\ \deq\ \exists y_i:
    \begin{conj}
      er_i(x,y_i)\\
      \forall var_{abs}, var_{abs}': J \land J' \implies ea(x)
    \end{conj}
  \]
  Intuitively, condition (\ref{eq:ref-fair}) requires to prove that
  any state in a run of the refinement that corresponds to a state
  satisfying the abstract fairness condition of event $ea(x)$ is
  followed either by the occurrence of one of the refining actions or
  by a state that no longer satisfies the fairness condition. Although
  the formal statement is somewhat technical, the abstract-level
  fairness condition is conveniently represented as a concrete-level
  ``leads to'' formula that can be established using the proof system
  of Fig.~\ref{fig:rules}. In particular, any fairness conditions of
  the implementation may be used, as well as induction over
  well-founded orderings. In this way, a specifier has much more
  freedom in justifying a refinement than with the more traditional
  verification conditions
  of~\cite{abrial:dynamic,back:refinement-fair}.
\end{itemize}

Using a standard simulation argument that critically relies on the
possibility of stuttering in the definition of runs of event systems,
one obtains the following correctness theorem: every run of the
refined event system $Ref$ corresponds to a run of the abstract event
system $Abs$, modulo the gluing invariant.

\begin{theorem}\label{thm:refine-runs}
  Assume that $Ref$ is a refinement of $Abs$ with respect to the
  gluing invariant $J$ and that $\sigma = s_0 s_1 \ldots$ is a run of
  $Ref$. Then there is a run $\tau = t_0 t_1 \ldots$ of $Abs$ such
  that $J$ holds at the joint valuations obtained from $s_i$ and
  $t_i$, for all $i \in \nat$.
\end{theorem}

As a consequence, temporal logic properties can be transferred from an
abstract event system $Abs$ to its refinement $Ref$ modulo the gluing
invariant $J$. Formally, this is asserted by the following corollary.

\begin{corollary}\label{thm:refine-props}
  Assume that $Ref$ is a refinement of $Abs$ with respect to the
  gluing invariant $J$ and that $\sigma = s_0 s_1 \ldots$ is a run of
  $Ref$. If $Abs \models \varphi$ then $Ref \models \O{\varphi}$ where
  $\O{\varphi}$ is obtained from $\varphi$ by replacing every positive
  occurrence of a non-temporal formula $A$ by $\exists var_{abs}: J
  \land A$ and every negative occurrence by $\forall var_{abs}: J
  \implies A$.
\end{corollary}

\subsection{Refinement preserving access control}
\label{sec:refine-secure}

Let us now consider how refinement interacts with access control
policies. Assume that event system $Ref$ is a refinement of $Abs$ with
respect to the gluing invariant $J$. Also, assume that $Abs$ was known
to implement certain permissions, prohibitions, obligations or user
rights concerning an abstract event $ea(x)$.

For permissions $perm_{ea}(x)$ and prohibitions $proh_{ea}(x)$, the
conditions (\ref{eq:permission}) and (\ref{eq:prohibition}) ensure
that $perm_{ea}(x)$ and $\lnot proh_{ea}(x)$ hold whenever event
$ea(x)$ occurs in a run of $Abs$. Any concrete-level event $er(x,y)$
refining $ea(x)$ has to satisfy condition (\ref{eq:ref-event}). Using
the definition of feasibility (\ref{eq:fis}) and first-order logic, it
follows that $\O{perm_{ea}(x)}$ and $\O{\lnot proh_{ea}(x)}$ hold
whenever event $er(x)$ occurs in a run of $Ref$. This is the best
preservation result we can hope for in such a general discussion of
refinement modulo a gluing invariant; for most practical choices of
$J$ these formulas will imply that the abstract-level permissions and
prohibitions are preserved in the refined system.

Similarly, obligations have been interpreted as liveness properties,
represented by the temporal logic formulas (\ref{eq:strict-obl}) or
(\ref{eq:weak-obl}). Corollary~\ref{thm:refine-props} implies that a
similar ``leads to'' formula is true of the refined model, again
modulo translation along the gluing invariant. Therefore, obligations
are preserved in the same sense as permissions and prohibitions.

These preservation results are not really surprising: we have
interpreted permissions, prohibitions, and obligations as (safety or
liveness) properties of runs, and the refinement notion of event
systems are defined in such a way that properties of runs are
preserved. However, we have also considered user rights, which were
interpreted as branching properties in Sect.~\ref{sec:user-rights},
and refinement of event systems does not necessarily preserve
branching behavior.

For a concrete example, consider a proposed refinement of the event
$extraPayback$ shown in Fig.~\ref{fig:bank-refined}. Instead of an
atomic event modeling an extra payment, the refinement introduces a
protocol: the client has to apply for making an extra payment (event
$askPayback$), and this application can be approved or rejected by the
bank, depending on the situation of the loan. The refinement is
acceptable according to the conditions (\ref{eq:ref-event}) and
(\ref{eq:ref-skip}) because $approvePayback$ refines the abstract
event $extraPayback$ whereas the events $askPayback$ and
$rejectPayback$ are unobservable at the abstract level. However, the
refinement does not literally preserve the user right
\[
\begin{noj}
  \event\ extraPayBack(l,amt)\\
  \allow\ l \in loans \land amt \in \rat \land amt + extra(l) \leq maxExtra(l).
\end{noj}
\]
considered in Sect.~\ref{sec:user-rights}: the concrete-level event
$approvePayback$ requires the precondition $(l \mapsto amt) \in
askExtra$, which is not implied by the predicate specifying the user
right. Preservation of user rights thus requires extra consideration.

\begin{figure}[tp]
  \centering
  \begin{tabular}{l}
    \event\ $askPayback(l,amt) =$\\
    \quad$\begin{conj}
      l \in loans \land amt \in \rat\\
      askExtra' = askExtra \cup \{l \mapsto amt\}\\
      loans' = loans \land clt'=clt \land due'=due \land rate' = rate\\
      maxExtra'=maxExtra \land extra'=extra
    \end{conj}$\\
    \allow\ $l \in loans \land amt \in \rat$\\[1mm]
    \event\ $approvePayback(l,amt) =$\\
    \quad$\begin{conj}
      (l \mapsto amt) \in askExtra \land amt + extra(l) \leq maxExtra(l)\\
      due' = due \oplus \{l \mapsto due(l) - amt\}
      \land extra' = extra \oplus \{l \mapsto extra(l) + amt\}\\
      askExtra' = askExtra \setminus \{l \mapsto amt\}\\
      loans' = loans \land clt' = clt \land rate' = rate
      \land maxExtra' = maxExtra
    \end{conj}$\\
    \fairness\ $(l \mapsto amt) \in askExtra \land amt + extra(l) \leq maxExtra(l)$\\[1mm]
    \event\ $rejectPayback(l,amt) =$\\
    \quad$\begin{conj}
      (l \mapsto amt) \in askExtra \land amt + extra(l) > maxExtra(l)\\
      askExtra' = askExtra \setminus \{l \mapsto amt\}\\
      due' = due \land extra' = extra\\
      loans' = loans \land clt' = clt \land rate' = rate
      \land maxExtra' = maxExtra
    \end{conj}$\\
  \end{tabular}
  \caption{Refining event $extraPayback$.}
  \label{fig:bank-refined}
\end{figure}

A first idea would be to impose the condition
\begin{equation}
  \label{eq:ref-right1}
  Hyp\ \models\ Inv_{ref} \land \O{right_{ea}(x)} \implies
  (\exists y_1: \fis er_1(x,y_1)) \lor\ldots\lor
  (\exists y_n: \fis er_n(x,y_n))
\end{equation}
where again $er_1(x,y_1)$, \ldots, $er_n(x,y_n)$ are the
concrete-level events corresponding to the abstract event $ea$.
Although condition (\ref{eq:ref-right1}) obviously preserves user
rights, it would rule out the refinement of
Fig.~\ref{fig:bank-refined}. More generally, this condition appears
too strong to us, when the concrete model refines the grain of
atomicity. Recall that a single abstract-level event $ea$ can be
implemented in the refinement by a sequence of concrete-level events
all but the last of which are invisible at the abstract level. The
final event $er$ refining the abstract event $ea$ need not be
immediately feasible in the concrete model whenever $ea$ is, but it
requires preparation by the auxiliary events that are unobservable at
the abstract level. We therefore believe that a more useful condition
for refining user rights is to require a combination of concrete-level
user rights that ensure that the branch leading to $er$ can be started
and concrete-level obligations that ensure that $er$ will then occur
eventually.

Formally, assume that the abstract system specification contains an
event $ea(x)$ for which we wish to ensure a user right via predicate
$right_{ea}(x)$. Also assume that $ea(x)$ is refined by the
concrete-level events $er_1(x,y_1)$, \ldots, $er_n(x,y_n)$. We then
require the event system $Ref$ to contain events $ei_1(x,z_1)$,
\ldots, $ei_m(x,z_m)$ with user rights specified by
$right_{ei_j}(x,z_j)$ such that
\begin{eqnarray}
  \label{eq:ref-right-init}
  \O{right_{ea}(x)} & \implies & 
  (\exists z_1: right_{ei_1}(x,z_1)) \lor\ldots\lor 
  (\exists z_m: right_{ei_m}(x,z_m)) \qquad\text{and}\\
  \label{eq:ref-right-term}
  ei_j(x,z_j) & \leadsto &
  \lnot \O{right_{ea}(x)} \lor
  (\exists y_1: er_1(x,y_1)) \lor\ldots\lor (\exists y_n: er_n(x,y_n))
\end{eqnarray}

Condition (\ref{eq:ref-right-term}) applies for all $j=1,\ldots,m$;
the disjunct $\lnot \O{right_{ea}(x)}$ on the right-hand side of
(\ref{eq:ref-right-term}) corresponds to a weak interpretation of
obligations.

The above conditions, together with the interpretations of the user
rights for the refined specification, imply that whenever the
translated abstract user right holds at some point during a
concrete-level run, the user has a concrete-level right to start a
branch which will eventually lead to the occurrence of an event
refining the original event $ea(x)$ provided the abstract-level right
persists. For example, the abstract-level right may cease due to the
concurrent exercise of another right.

Back to the example of Fig.~\ref{fig:bank-refined}, we claim that this
refinement respects the abstract-level user right because it satisfies
the conditions (\ref{eq:ref-right-init}) and
(\ref{eq:ref-right-term}). We assume that the gluing invariant
contains the conjunct
\[
  askExtra \subseteq loans \times \rat
\]
that asserts the ``type correctness'' of the new variable $askExtra$.
We choose $askPayback$ for the auxiliary event $ei$, and condition
(\ref{eq:ref-right-init}) boils down to proving
\[
  l \in loans \land amt \in \rat \land amt+extra(l) \leq maxExtra(l)
  \implies l \in loans \land amt \in \rat
\]
which is trivial. On the other hand, condition
(\ref{eq:ref-right-term}) requires us to show
\[
\begin{noj}
  askPayback(l,amt) \leadsto
  \begin{disj}
    \lnot (l \in loans \land amt \in \rat \land amt+extra(l) \leq maxExtra(l))\\
    grantPayback(l,amt)
  \end{disj}
\end{noj}
\]
and this condition is ensured by the fairness condition for event
$approvePayback$. Note that although the abstract user right is
preserved, the client cannot cheat on the bank by demanding two extra
payments that together would exceed the allowed limit: although a
client may always ask for an extra payment (including in the time
between applying for a payment and the approval or rejection by the
bank), the bank's obligation to approve extra payments ceases when the
limit has been reached, so it is free to reject a second application
for extra payments. This is just what the abstract user right of
Sect.~\ref{sec:user-rights} required.

\section{Conclusion}
\label{sec:conclusion}

Event systems are a convenient and widely accepted framework for
modeling information systems. In particular, properties of their runs
can be derived using well-known rules, and refinement concepts for
event systems are well established. In this paper, we have considered
annotating event systems with clauses to specify access control
properties, thereby implementing a given security policy. Existing,
declarative languages for describing access control such as OrBAC
identify the static structure of an information system, including the
subjects, the objects, and the activities, and then spell out the
conditions under which activities may, must, or must not be performed.
In this paper, we have interpreted such policies within a formal
system model based on event systems, and have proposed proof rules for
verifying that a system implements a security policy. We have
considered permissions and prohibitions, which are the most frequent
annotations in practice, and which can be interpreted as safety
properties of system runs. We have proposed to interpret obligations
as liveness properties, and have therefore used a simple temporal
logic to formulate these as properties of event systems. As a fourth
category of primitives, we have considered user rights, which can be
interpreted as elementary branching properties of systems.

Development methods based on stepwise refinement have traditionnally
been associated with event systems. They allow a developer to justify
a system as a result of a sequence of models that introduce more and
more details in the representation of systems, as well as their
correctness properties.  The cornerstone of refinement is the
preservation of properties that have been established for abstract
models. Standard refinement concepts preserve traces of models, and
this ensures preservation of permissions, prohibitions, and
obligations across refinements. Branching properties, including user
rights, are not automatically preserved, and we have proposed
additional conditions that rely on a combination of concrete-level
rights and obligations.

More experience will be necessary to evaluate whether our notions are
useful and feasibility in practice. It would also be helpful to have
an integrated tool environment for combining event system descriptions
and access control specifications. On a more conceptual level, it will
be interesting to study the possibility of synthesizing security
monitors that enforce a security policy (that could possibly even vary
during runtime) over a fixed underlying information system.

\bibliographystyle{plain}
\bibliography{mybib}

\begin{thebibliography}{10}

\bibitem{abadi:existence}
Mart\'{\i}n Abadi and Leslie Lamport.
\newblock The existence of refinement mappings.
\newblock {\em Theoretical Computer Science}, 81(2):253--284, May 1991.

\bibitem{abrial:b-book}
J.-R. Abrial.
\newblock {\em The B-Book: Assigning Programs to Meanings}.
\newblock Cambridge University Press, 1996.

\bibitem{abrial:event}
Jean-Raymond Abrial.
\newblock {B\textsuperscript{\#}}: Toward a synthesis between {Z} and {B}.
\newblock In D.~Bert, J.~P. Bowen, S.~King, and M.~A. Wald{\'e}n, editors, {\em
  Formal Specification and Development in Z and B (ZB~2003)}, volume 2651 of
  {\em Lecture Notes in Computer Science}, pages 168--177. Springer Verlag,
  2003.

\bibitem{abrial:refinement}
Jean-Raymond Abrial, Dominique Cansell, and Dominique M{\'e}ry.
\newblock {Refinement and Reachability in Event\_B}.
\newblock In H.~Treharne, S.~King, and M.~Henson, editors, {\em {Formal
  Specification and Development in Z and B (ZB~2005)}}, volume 3455 of {\em
  Lecture Notes in Computer Science}, pages 222--241, Guilford, UK, 2005.
  Springer Verlag.

\bibitem{abrial:dynamic}
Jean-Raymond Abrial and Louis Mussat.
\newblock Introducing dynamic constraints in {B}.
\newblock In D.~Bert, editor, {\em Advances in the Development and the Use of
  the {B} Method}, volume 1393 of {\em Lecture Notes in Computer Science},
  pages 83--128. Springer Verlag, 1998.

\bibitem{arnold:games-partial}
Andr\'{e} Arnold, Aymeric Vincent, and Igor Walukiewicz.
\newblock Games for synthesis of controlers with partial observation.
\newblock {\em Theoretical Computer Science}, 303(1):7--34, 2003.

\bibitem{back:refinement}
R.~Back and J.~von Wright.
\newblock {\em Refinement calculus---A systematic introduction}.
\newblock Springer-Verlag, 1998.

\bibitem{back:refinement-fair}
Ralph-Johan Back and Qiwen Xu.
\newblock Refinement of fair action systems.
\newblock {\em Acta Informatica}, 35:131--165, 1998.

\bibitem{benferhat:stratification}
S.~Benferhat, R.~El Baida, and F.~Cuppens.
\newblock A stratification-based approach for handling conflicts in access
  control.
\newblock In {\em 8th ACM Symp. Access Control Models and Technologies
  (SACMAT'03)}, 2003.

\bibitem{bertino:logical}
Elisa Bertino, Barbara Catania, Elena Ferrari, and Paolo Perlasca.
\newblock A logical framework for reasoning about access control models.
\newblock {\em ACM Trans. Inf. Syst. Secur.}, 6(1):71--127, 2003.

\bibitem{cuppens:misconfiguration}
F.~Cuppens, N.~Cuppens-Boulahia, and J.~Garcia.
\newblock Misconfiguration management of network security components.
\newblock In {\em IASTED International Conference on Communication, Network,
  and Information Security (CNIS 2005)}, 2005.

\bibitem{cuppens:inheritance}
F.~Cuppens, N.~Cuppens-Boulahia, and A.~Mi\`{e}ge.
\newblock Inheritance hierarchies in the or-bac model and application in a
  network environment.
\newblock In A.~Sabelfeld, editor, {\em Proc. Foundations of Computer Security
  (FCS04)}, pages 41--60, Turku,Finland, 2004. Turku Center for Computer
  Science, Report G-31.

\bibitem{guelev:access-control}
Dimitar~P. Guelev, Mark Ryan, and Pierre-Yves Schobbens.
\newblock Model-checking access control policies.
\newblock In {\em Proc. 7th Intl. Conf. Information Security (ISC 2004)},
  volume 3225 of {\em Lecture Notes in Computer Science}, pages 219--230, Palo
  Alto, CA, 2004. Springer.

\bibitem{hilpinen:deontic}
Risto Hilpinen, editor.
\newblock {\em New Studies in Deontic Logic: Norms, Actions, and the
  Foundations of Ethics}, volume 152 of {\em Synthese Library}.
\newblock D. Reidel, Dordrecht, Holland, 1981.

\bibitem{juerjens:secrecy}
J.~{J}{\"{u}}{r}jens.
\newblock Secrecy-preserving refinement.
\newblock In J.~Fiadeiro and P.~Zave, editors, {\em Intl. Symp. Formal Methods
  Europe (FME 2001)}, volume 2021 of {\em Lecture Notes in Computer Science},
  pages 135--152. Springer, 2001.

\bibitem{juerjens:umlsec}
Jan J\"urjens.
\newblock {\em Secure Systems Development with UML}.
\newblock Springer, 2004.

\bibitem{abou:orbac}
A.~Abou~El Kalam, R.~El Baida, P.~Balbiani, S.~Benferhat, F.~Cuppens,
  Y.~Deswarte, A.~Mi\`{e}ge, C.~Saurel, and G.~Trouessin.
\newblock Organization based access control.
\newblock In {\em 4th Intl. IEEE Workshop Policies for Distributed Systems and
  Networks}, pages 120--131, Como, Italy, 2003. IEEE Press.

\bibitem{koch:access-control}
Manuel Koch, Luigi~V. Mancini, and Francesco Parisi-Presicce.
\newblock Graph-based specification of access control policies.
\newblock {\em J. Comput. Syst. Sci.}, 71(1):1--33, 2005.

\bibitem{lamport:tla}
Leslie Lamport.
\newblock The {T}emporal {L}ogic of {A}ctions.
\newblock {\em {ACM} Transactions on Programming Languages and Systems},
  16(3):872--923, May 1994.

\bibitem{meyer:deontic}
J.-J. Meyer and R.~Wieringa, editors.
\newblock {\em Deontic Logic in Computer Science}.
\newblock Wiley, 1993.

\bibitem{misra:logic}
Jayadev Misra.
\newblock A logic for concurrent programming: Safety and progress.
\newblock {\em Journal of Computer and Software Engineering}, 3(2):239--300,
  1995.

\bibitem{sandhu:rbac}
Ravi Sandhu, Edward Coyne, Hal Feinstein, and Charles Youman.
\newblock Role-based access control models.
\newblock {\em IEEE Computer}, 29(2):38--47, 1996.

\bibitem{zhang:evaluating}
Nan Zhang, Mark Ryan, and Dimitar~P. Guelev.
\newblock Evaluating access control policies through model checking.
\newblock In Jianying Zhou, Javier Lopez, Robert~H. Deng, and Feng Bao,
  editors, {\em Proc. 8th Intl. Conf. Information Security (ISC 2005)}, volume
  3650 of {\em Lecture Notes in Computer Science}, pages 446--460, Singapore,
  2005. Springer.

\end{thebibliography}

\end{document}